# Regression for Citation Data: An Evaluation of Different Methods[1]

Mike Thelwall, Paul Wilson

**Citations are increasingly used for research evaluations. It is therefore important to identify factors affecting citation scores that are unrelated to scholarly quality or usefulness so that these can be taken into account. Regression is the most powerful statistical technique to identify these factors and hence it is important to identify the best regression strategy for citation data. Citation counts tend to follow a discrete lognormal distribution and, in the absence of alternatives, have been investigated with negative binomial regression. Using simulated discrete lognormal data (continuous lognormal data rounded to the nearest integer) this article shows that a better strategy is to add one to the citations, take their log and then use the general linear (ordinary least squares) model for regression (e.g., multiple linear regression, ANOVA), or to use the generalized linear model without the log. Reasonable results can also be obtained if all the zero citations are discarded, the log is taken of the remaining citation counts and then the general linear model is used, or if the generalized linear model is used with the continuous lognormal distribution. Similar approaches are recommended for altmetric data, if it proves to be lognormally distributed.**

**Keywords:** scientometrics, informetrics, altmetrics, citation distributions, lognormal, powerlaw, regression

## Introduction

The use of performance monitoring for university research has increased over the past few decades. This is most evident in national research evaluation exercises, such of those in the UK (Mryglod, Kenna, Holovatch, & Berche, 2013), Australia (ARC, 2014), New Zealand (Anderson, Smart, & Tressler, 2013) and Italy (Abramo, D'Angelo, & Di Costa, 2011). This climate not only affects the allocation of research funding in many cases but can also change the behaviour of individual researchers as they come to terms with the assessment system (Butler, 2003). Although the most important performance monitoring exercises often rely on peer review, both the UK (REF, 2013) and Australia (ARC, 2014) consider citations for some subject areas, and there are advocates of increasing use of citations for some types of science when the results correlate because citation metrics are much cheaper than peer review (Abramo, Cicero, & D'Angelo, 2013; Franceschet, & Costantini, 2011; Mryglod et al., 2013), although no simple method is likely to work (Butler & Mcallister, 2011). In addition, citations are used for formal and informal evaluations of academics (Cole, 2000) and the Journal Impact Factor (JIF) is a widely recognised and used citation metric.

The range of metrics of relevance to science has recently increased with the emergence of webometrics (Almind & Ingwersen, 1997), which includes a range of new impact indicators derived from the web (Kousha & Thelwall, 2014) and altmetrics (Priem, Taraborelli, Groth, & Neylon, 2010), which incorporate many attention and impact indicators derived from social web sites (Priem, 2014). Altmetrics seem particularly promising to help researchers to identify recently-published articles that have attracted a lot

---

of attention (Adie & Roe, 2013) and to give evidence of non-standard impacts of research that can be added to CVs (ACUMEN, 2014; Piwowar & Priem, 2013). Statistical analyses of some of these have started to generate new insights into how science works (Mohammadi & Thelwall, 2014; Thelwall & Maflahi, in press) and the types of research impacts that are not recognised by traditional citation counts (Mohammadi & Thelwall, 2013).

Because of the many uses of citations within science, it is important to understand as much as possible about why they are created and why one article, researcher or research group may be more cited than another. Whilst citations may be given to acknowledge relevant previous work (Merton, 1973), they can also be used to criticise or argue against it (MacRoberts & MacRoberts, 1996) and so citations are not universally positive. Moreover, citations do not appear to be chosen in a dispassionate, unbiased way. For example, researchers in some fields tend to cite papers written in the same language (Yitzhaki, 1998), highly relevant citations may be systematically ignored (McCain, 2012) and fame seems also to attract citations (Merton, 1968). There are also field differences in the typical number of citations received by papers (e.g., Waltman, van Eck, van Leeuwen, Visser, & van Raan, 2011) and review articles are more likely to be highly cited than other articles (e.g., Aksnes, 2003). From a different perspective, factors that associate with highly cited papers, such as collaboration, internationality and referencing patterns (Didegah & Thelwall, 2013; Sooryamoorthy, 2009), are important because they may push researchers and funders towards more successful types of research.

Although some of the factors affecting citations discussed above have been discovered using qualitative methods, such as interviews with authors, statistical methods are needed to identify the magnitude of factors and to detect whether they apply in particular cases. The simplest approach is probably to compare the average number of citations (or citation-based indicators) for one set of papers against that of another to see which is higher (van Raan, 1998). Another strategy is to assess whether citation counts correlate significantly against another metric that is hypothesised to be related (Shema, Bar-Ilan, & Thelwall, 2014). The most discriminating methods used so far are regression-based because they allow the effects of multiple variables to be examined simultaneously. In particular, regression guards against one factor being identified as significant (e.g., international collaboration) when another factor (e.g., collaboration with the USA) is underlying cause of higher (or lower) citations.

There is no consensus about the best regression method for citation data. Methods used so far include ordinary least squares linear regression (Aksnes, Rørstad, Piro, & Sivertsen, 2013; Dragos & Dragos, 2014 [citations per publication used as the dependant variable]; Foo & Tan, 2014; He, 2009; Mavros, Bardakas, Rafailidis et al., 2013; Rigby, 2013 [adding 1 to citations, dividing by a time normalised value and taking their log]; Tang, 2013 [adding 1 to citations and taking their log]; Stewart, 1983), logistic regression (Baldi, 1998; Bornmann, & Williams, 2013; Kutlar, Kabasakal, & Ekici, 2013; Sin, 2011; Willis, Bahler, Neuberger, & Dahm, 2011; Xia & Nakanishi, 2012; Yu, Yu, & Wang, 2014), a distribution-free regression method (Peters & van Raan, 1994), multinomial logistic regression (Baumgartner & Leydesdorff, 2014) and negative binomial regression (Chen, 2012; Didegah & Thelwall, 2013ab; McDonald, 2007; Thelwall & Maflahi, in press [for altmetrics]; Walters, 2006; Yoshikane, 2013 [for patent citations]).

The typical distribution of citations is highly skewed (Price, 1965; Seglen, 1992), so that tests based upon the normal distribution (e.g., ordinary least squares regression) are not appropriate if the data is raw citation counts. Logistic regression can avoid this issue by predicting membership of the highly cited group of papers rather than directly predicting citations. Whilst negative binomial regression can cope with skewed data and is designed for discrete numbers (Hilbe, 2011), the most appropriate distribution for citations to a collection of articles from a single subject and year seems to be the discrete lognormal distribution (Evans, Hopkins, & Kaube, 2012; Radicchi, Fortunato, & Castellano, 2008; Thelwall & Wilson, 2014) and the hooked power law is also a reasonable choice (Thelwall & Wilson, 2014). Although many articles suggest a power law for the tail of citation distributions (e.g., Yao, Peng, Zhang, & Xu, 2014) this is not helpful for statistical analyses that need to include all cited articles and is broadly consistent with a lognormal distribution for all articles, although small discrepancies may be revealed by fine-grained analyses (Golosovsky & Solomon, 2012). Citations to articles from individual journals almost always conform to a lognormal distribution (Stringer, Sales-Pardo, & Amaral, 2010), as do some other citation-based indicators also follow a lognormal distribution (e.g., generalised h-indices: Wu, 2013). Although it has not been fully tested, it seems likely that most sets of articles from a specific time window will approximately follow a discrete lognormal distribution, unless the time window is too long or very recent. Hence it is not clear that negative binomial regression is optimal when the dependant variable is citation counts.

Neither the discrete lognormal or the hooked power law distributions have been used for regression because it seems that no software exists for this. An alternative strategy would be to take the log of the citations and then use the general linear model to analyse them with the assumption that the logged citations would be normally distributed (the general linear model assumes that the error terms or residuals are normally distributed). Although the log of the continuous version of the lognormal distribution is a perfect normal distribution, the same is not true for the discrete lognormal distribution and so it is not obvious that this will work. Moreover, the use of log transformation for citation data has been argued against for classification purposes because of the variance reduction that it introduces (Leydesdorff & Bensman, 2006), but this is not evidence that it will not work for regression. This article assesses both of these and the continuous normal distribution in order to identify the most powerful, regression-based approach for citations and similar data, such as altmetrics. The results will help to ensure that future statistical analyses of the factors affecting citation counts are as powerful and reliable as possible.

# Citation distributions and statistical tests

Statistical regression models attempt to identify the relationship between a dependant variable (citations in this case) and one or more independent variables (e.g., the field in which an article is published or the number of authors). The relationship is typically identified by an algorithm that estimates parameters for the independent variables in order to generate a model that, for any given set of their values, predicts an expected value for the dependant variable. For example, if citations were regressed against scientific field, then, for any given field, the equation would predict the expected number of citations for that field. Moreover, tests associated with the regression could confirm or reject the hypothesis that two or more fields have different means. The strategy for fitting the model depends upon the distribution of the error terms, which are the difference between the predicted values and the observed values for the data. If these error terms are normally distributed then the regression can use the general linear (ordinary least squares) model (Garson, 2012). If the error terms are not normally distributed, however, then using the general linear model can give misleading results. To illustrate this, since citations are highly skewed, if the general linear model were to be used to fit an equation then the estimates made by the equation would be greatly influenced by the few highly cited articles because the normal distribution assumption would lead to a prediction of a higher estimated mean to reduce the otherwise extremely low chance of the high citation. In other words a citation that is high enough to be an outlier for the normal distribution could still be within the expected range of values for the lognormal distribution.

Switching to the generalized linear model (Dobson & Barnett, 2008) can resolve the problem of error terms not being normally distributed when fitting the general linear model. This allows the distribution of error terms to be specified as something other than normal, such as negative binomial, Poisson or lognormal, and fits the regression equation using maximum likelihood estimation (Millar, 2011). For example, if the distribution is specified as negative binomial, which is skewed, then the regression equation fitted to a citation distribution will be much less influenced by high citation scores than in the case of the general linear model, because high values are more likely for the negative binomial distribution. Hence negative binomial regression (Hilbe, 2011) seems to be a better choice than does general ordinary least squares linear regression.

The negative binomial model assumes that the error terms (i.e., residuals) follow a negative binomial distribution. Although this is a skewed distribution, it is not the same as the lognormal distribution and hence is unlikely to fit lognormal data well, although it is not clear whether the difference between the two approaches would be significant in practice. Here, a discrete version of the lognormal distribution would ideally be needed to model the error terms for the generalized linear model. Unfortunately, however, whilst many different versions of the generalized linear model are available in statistical software, such as R, SPSS and STATA, at the time of writing (June 2014) none includes either the discrete lognormal distribution and so this is not a practical option. Note that although R includes the Poisson lognormal distribution (Aitchison & Ho, 1989), this is different from the discrete lognormal distribution. The GAMLSS package in R can be used to fit models based upon the continuous lognormal distribution (Stasinopoulos & Rigby, 2007) with the generalized linear model and so this is another logical alternative.

An alternative to generalized linear regression with a lognormal error function is to take the log of the discrete lognormal data and analyse it with the general linear model with the assumption that the error distribution, despite being derived from a discrete version of the lognormal distribution, will be close enough to normal for models to be effectively fitted. This is plausible because large values have the most influence on model fitting and the log function eliminates these. Nevertheless, there is a problem in that the log function is undefined for zero citations. To circumvent this, either all zero citations should be removed, or an offset of 1 should be added to all citation counts (a common statistical method, e.g., Tabachnick & Fidell, 2001, p. 81) to allow the logs to be calculated (Tang, 2013). Neither of these is an ideal solution since both introduce a systematic source of bias into the data. The first method (removing all zero citations) seems to be the most problematic, however, since it involves truncating one side of the distribution, which will bias the regression towards estimating higher citation means and may affect factors associating with lower citation scores more than factors associating with higher citation scores, reducing the power of the test. In addition, the truncation reduces the sample size and hence also the power of the test.

Finally, it is possible, although it seems unlikely, that citations do not follow a lognormal distribution but instead the citations to a collection of articles follow another distribution but with factors that combine the other distribution in a way that gives a combined distribution that is lognormal. Whilst theoretically possible, this is unlikely given the range of different sets of articles that have been tested and found to be approximately discrete lognormal (Thelwall & Wilson, 2014).

## Research question

In the absence of software that uses a discrete version of the lognormal distribution in the generalized linear model, this article assesses three logical alternatives, driven by the following research question: Which of the following is the most appropriate for discrete lognormal data in the sense that it does not give an inflated chance of false positive results whilst being powerful enough to distinguish minor factors in the data?

1. Negative binomial regression for the raw data.
2. Regression with the generalized linear model with a continuous lognormal distribution for the error terms after removing zeroes.
3. Regression with the general linear model on the log of the raw data after removing zeroes.
4. Regression with the generalized linear model with a continuous lognormal distribution for the error terms after adding 1 to the observations.
5. Regression with the general linear model on the log of the raw data after adding 1 to the observations.

## Methods

In order to address the research question, the three regression approaches must be applied to citation-like data with known relationships. This can be achieved by tests on simulated discrete lognormal data with and without relationships. Although there are infinitely many combinations of dependant variables that could be tested, the simplest case will give the clearest results and so only one relationship will be tested: that of a binary factor so that the two values of the factor correspond to different means for the lognormal distribution (but the same standard deviation).

Datasets were randomly created with citations, $C$, and a binary factor, $B$, taking two values, 0 and 1, so that the population distribution was approximately lognormal (due to discretization) with mean $\mu_B$ and standard deviation $\sigma$. For simplicity, half of each sample was taken for $B = 0$ and half for $B = 1$. A range of differences between $\mu_0$ and $\mu_1$ was tested to see how large the difference must be in order to reliably detect the binary factor. This is related to the power of the test. The following R commands were used for all of tests reported here.

```
glm.nb(C ~ f, data = all_data) #negative binomial
gamlss(C ~ f, family = LOGNO(), data = zero_truncated_data) #continuous lognormal
aov(Clog ~ f, data= zero_truncated_data) #ANOVA on log of C
gamlss(C_plus1 ~ f, family = LOGNO(), data = all_data) #continuous lognormal +1
aov(C_plus1_log ~ f, data= all_data)  #ANOVA on log of C +1
```

To test for false identification of factors, data was drawn as above except that $\mu_0 = \mu_1$. If, despite the identical means, a factor is detected by regression then this is a false positive result.

The test procedure was to generate 1000 random datasets for each set of parameters to be tested and to calculate the proportion of the datasets giving correct results for each of the above five models. This procedure was repeated for a variety of different sample sizes, means and standard deviations to reduce the chance of the results being specific to the set of parameters chosen.

## Results

The five approaches were first fitted to discrete lognormal data without any factors using 1000 simulations at each of a range of different sample sizes. As shown in Figure 1, when the log of the mean is 0.5 and the log of the standard deviation is 1, both ANOVA variants and both continuous lognormal models incorrectly identify factors at the approximately the rate corresponding to the level of significance used (0.05), which is the desired behaviour and suggests that the deviation from normality of the logged discrete data does not cause a problem for the tests. In contrast, and as suggested by the discussion above, negative binomial regression incorrectly identifies a factor at about triple the rate of that suggested by the confidence level. When the log of the standard deviation is increased to 2, the ANOVA and continuous lognormal models keep the same approximately 5% failure rate but negative binomial regression identifies non-existent factors about half of the time and is *less* successful as the sample size increases (Figure 2). The same patterns also occur for log means of 1 and 1.5 and so the results do not seem to be dependent on the selected mean[2].

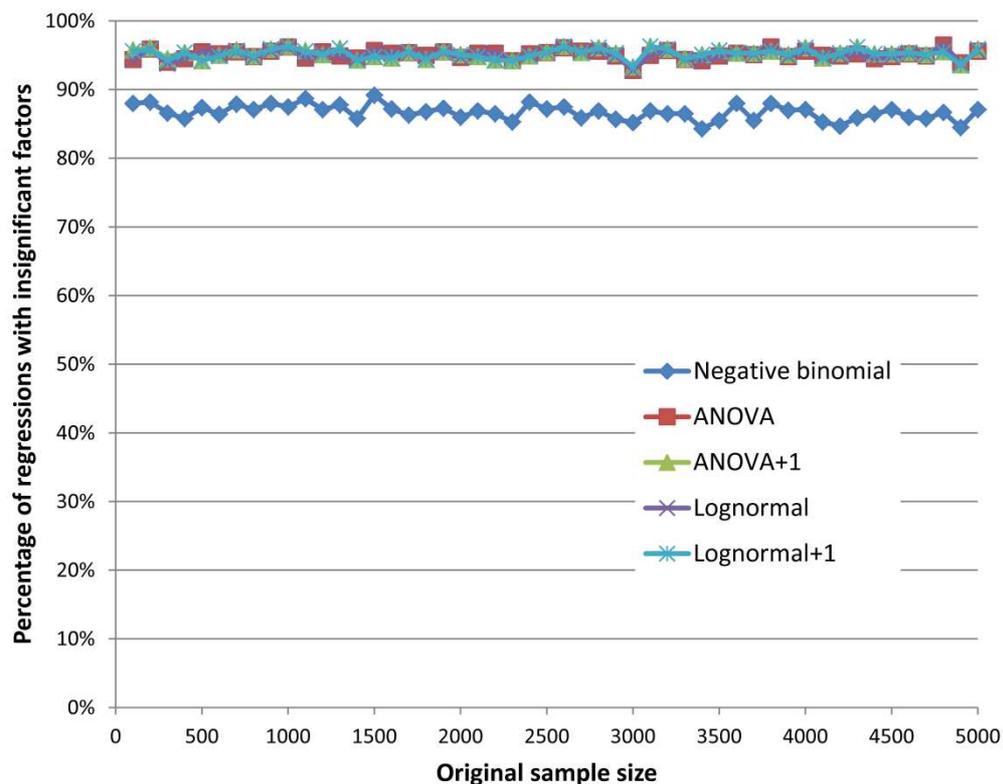

Fig. 1. Regressions against simulated discrete lognormal data (log mean: 0.5, log standard deviation: 1) at different sample sizes without factors (n=1000 iterations). All lines overlap except for the negative binomial line.

---

[2] The full code for the tests, together with all the data used and additional graphs (including any marked as "not shown" in the text) is available at
http://figshare.com/articles/Data_from_the_paper_Regression_for_Citation_Data_/1090729.

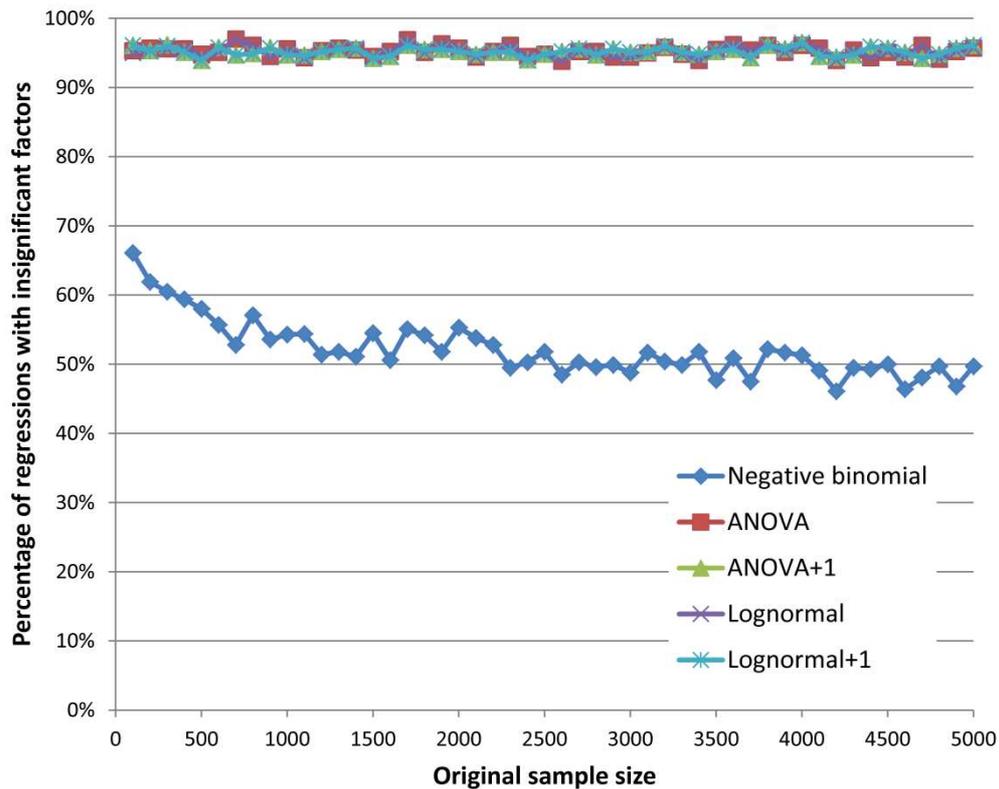

Fig. 2. Regressions against simulated discrete lognormal data (log mean: 0.5, log standard deviation: 2) at different sample sizes without factors (n=1000 iterations). All lines overlap except for the negative binomial line.

The regressions against simulated discrete lognormal data with a simple factor give information about the power of the tests to correctly identify factors. When the log of the means of the two values of the factor differ by only 0.05 (i.e., one log mean is 0.50 and the other is 0.55) all of the methods need larger sample sizes to increase their chance of detecting it (Figure 3). The most powerful of the three methods is negative binomial regression and the least powerful are the standard ANOVA and the standard continuous lognormal model. The apparent additional power of the negative binomial over that of the standard ANOVA and continuous lognormal is about the same as the difference in Figure 1, suggesting that the additional positives are due to the tendency to generate false positive results rather than it being a more powerful method. The additional power of the ANOVA+1 and lognormal+1 tests are presumably due to the increased sample size and the increased difference between the two factors caused by not deleting the zeroes. Although Figure 3 is based on a single mean and standard deviation, the pattern is almost identical for log means of 1 or 1.5, and the relationship between the methods is the same if the log standard deviation is doubled to 2 (Figure 4), although this makes the factor harder to detect.

      If the factor is easier to detect because there is a higher difference in log means (e.g., 0.1 rather than 0.05) then (Figure 5) then the advantage of the modified ANOVA and lognormal over the standard ANOVA and continuous lognormal is more pronounced.

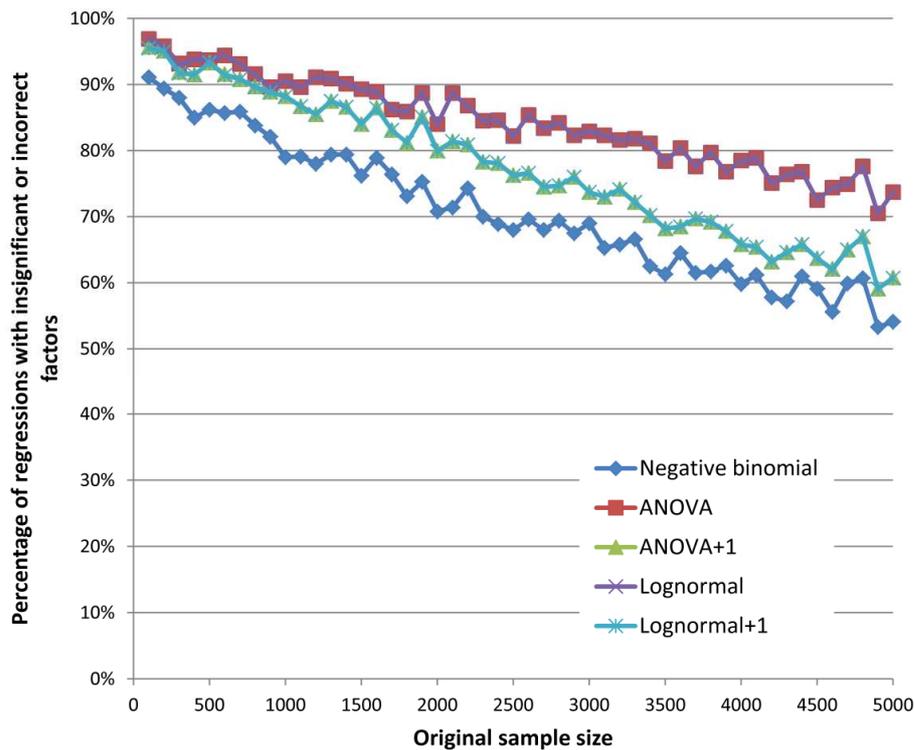

Fig. 3. Regressions against simulated discrete lognormal data (log standard deviation: 1) at different sample sizes with a binary factor with log means 0.5 and 0.55 (n=1000 iterations). The ANOVA and lognormal lines and the lognormal+1 and ANOVA+1 lines overlap.

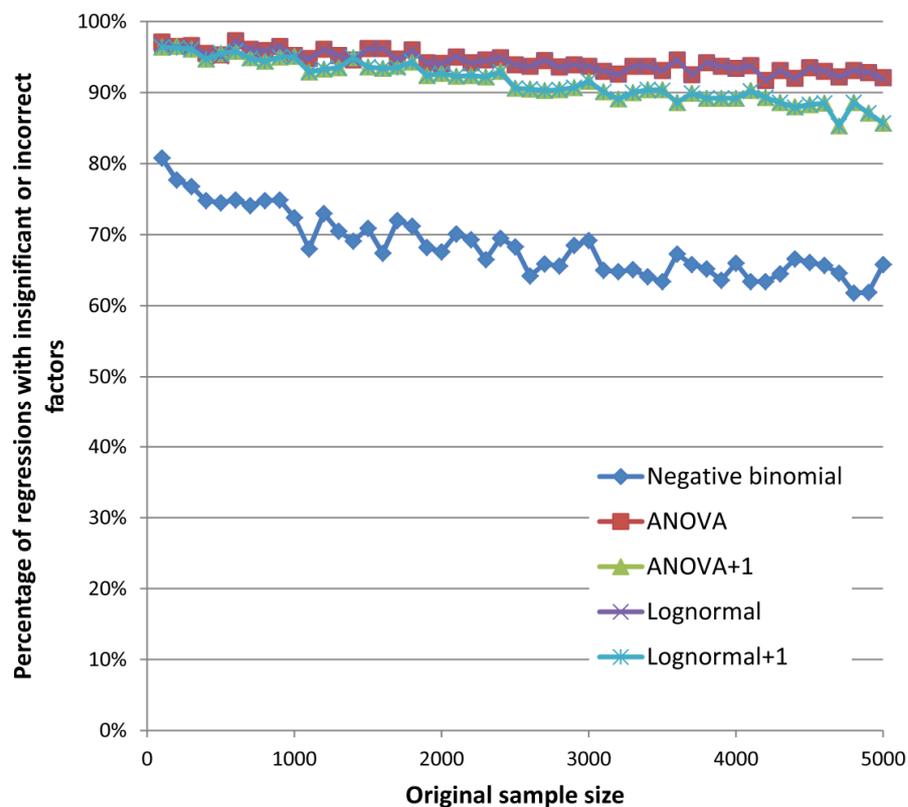

Fig. 4. Regressions against simulated discrete lognormal data (log standard deviation: 2) at different sample sizes with a binary factor with log means 0.5 and 0.55 (n=1000 iterations). The ANOVA and lognormal lines and the lognormal+1 and ANOVA+1 lines overlap.

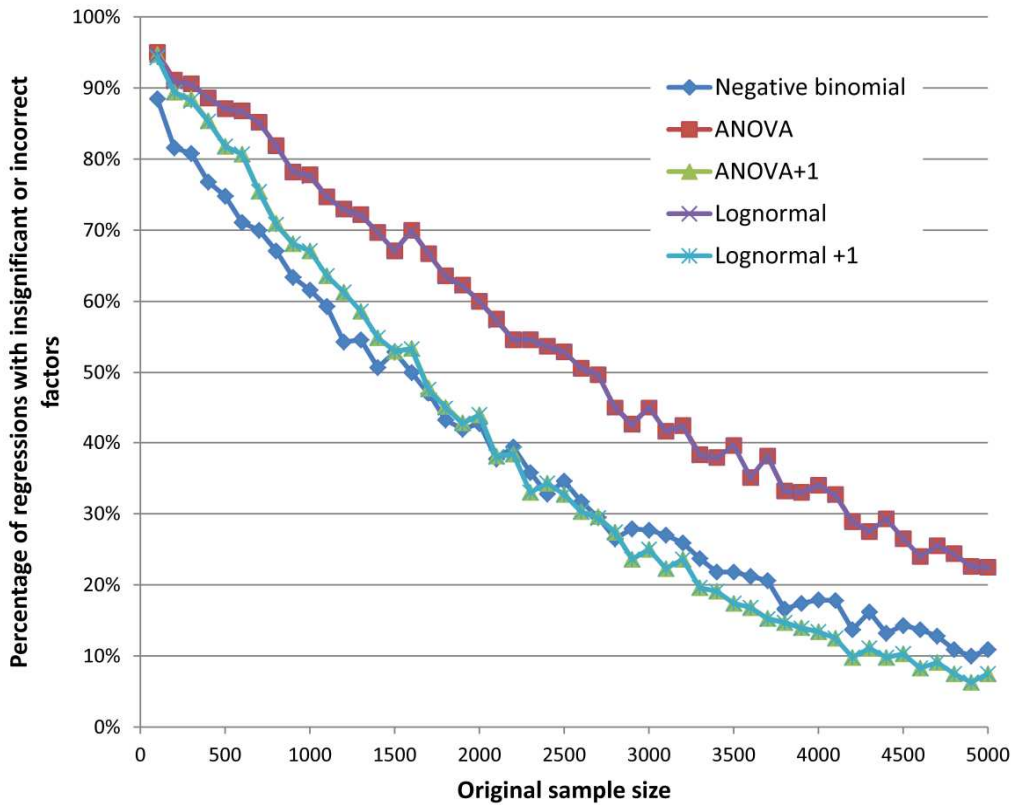

Fig. 5. Regressions against simulated discrete lognormal data (log standard deviation: 1) at different sample sizes with a binary factor with log means 0.5 and 0.6 (n=1000 iterations). The ANOVA and lognormal lines and the lognormal+1 and ANOVA+1 lines overlap.

# Discussion and conclusions

A limitation of the tests reported here is that they only deal with the simplest case of a single factor but it seems likely that the same conclusions would be drawn for more complex factors. Moreover, the unreliability of negative binomial regression for citations is probably overestimated by the simulation approach used here since real citation data will be bounded, especially if the data is from a small set of years that are not a long time in the past.

The results show that negative binomial regression applied to discrete lognormal data will identify non-existent factors in at a higher rate than expected by the significance level and so conclusions based upon negative binomial regression in such circumstances cannot be trusted. In contrast, the strategy of taking the logarithm of the lognormal data after discarding zeroes, and then applying the general linear model (ANOVA here, but other versions should also work) does not cause problems even though the resulting distribution is not normal. The situation for the generalized linear model with continuous lognormal error distribution is identical and the two approaches almost never give different results. The strategy of adding 1 to the data before taking its logarithm and then using the general linear model (as used by: Rigby, 2013; Tang, 2013) is similarly trustworthy and is equivalent to adding 1 to the data and then using the generalized linear model. The latter two are preferable, however, because they are more powerful in their ability to detect factors. This power to identify factors presumably derives from the larger sample size and increased difference between factors when the zeroes are not discarded.

In some cases, citation data has more uncited articles than would be expected by a standard statistical distribution (i.e., zero inflation). If this is suspected (e.g., by visual inspection of the distribution of citations for a particular dataset) then the zeroes can be removed and then either the log taken of the citations and the general linear model used (i.e., like the standard ANOVA model used above, without adding 1) or the generalized linear model used with the continuous lognormal distribution. Whilst these will be less powerful than using the full data set, they are unlikely to increase the chance of falsely identifying relationships and so are reasonable strategies.

If a discrete version of the lognormal distribution is created in the future for the generalized linear model then citation-like data could be regressed more precisely. If this happens then there would still be practical advantages with the above process (taking the log of the data and using the general linear model) because it seems intuitively unlikely that it would be substantially less powerful than the generalized linear model with the discrete lognormal distribution and it is a standard statistical test and hence there is fast and robust software for it that is widely understood and has many options.

The above strategies should be used for regressions of the mean (i.e., rather than logistic regression) for citation data that follows a discrete lognormal distribution. This seems likely to include most sets of articles with a common citation window but further studies are needed to identify general rules for when the lognormal distribution occurs. Nevertheless, researchers in the future should at least test for the lognormal distribution before using other models. The above strategies should also be considered for altmetrics, since they can correlate with citations (e.g., Li, Thelwall, & Giustini, 2012; Thelwall, Haustein, Larivière, & Sugimoto, 2013; Zahedi, Costas, & Wouters, in press) and hence it may have discrete lognormal distributions (e.g., article downloads: Yan & Gerstein, 2011), although preliminary studies of a range of altmetrics (data from: Thelwall et al., 2013) suggest that some may follow a power law instead[3]. Hence, the exact distributions of each one should be checked in case they follow different patterns.

---

[3] The full code for the tests, together with all the data used and additional graphs (including any marked as "not shown" in the text) is available at
http://figshare.com/articles/Data_from_the_paper_Regression_for_Citation_Data_/1090729.